\documentclass[nofootinbib,twocolumn,showpacs,preprintnumbers,pre,aps,superscriptaddress]{revtex4-1}

\usepackage[dvipdfmx]{graphicx}
\usepackage{bm}
\usepackage{amsmath}
\usepackage{amssymb}
\usepackage{color}

\begin{document}

\title{Lateral response of a layered material with interlayer friction}

\author{Tomoki Sasada}\email{These authors contributed equally to this work.}

\affiliation{
Department of Chemistry, Graduate School of Science,
Tokyo Metropolitan University, Tokyo 192-0397, Japan}

\author{Kento Yasuda}\email{These authors contributed equally to this work.}

\affiliation{
Research Institute for Mathematical Sciences, 
Kyoto University, Kyoto 606-8502, Japan}

\author{Yuto Hosaka}

\affiliation{
Max Planck Institute for Dynamics and Self-Organization (MPI DS), 
Am Fassberg 17, 37077 G\"{o}ttingen, Germany}

\author{Shigeyuki Komura}\email{komura@wiucas.ac.cn}

\affiliation{
Wenzhou Institute, University of Chinese Academy of Sciences, 
Wenzhou, Zhejiang 325001, China} 

\affiliation{
Oujiang Laboratory, Wenzhou, Zhejiang 325000, China}

\affiliation{
Department of Chemistry, Graduate School of Science,
Tokyo Metropolitan University, Tokyo 192-0397, Japan}


\begin{abstract}
We investigate the mechanical properties of a layered material with interlayer friction.
We propose a model that contains lateral elasticity and interlayer friction to obtain 
the response function both in the Fourier and real spaces. 
By investigating how the internal deformation is laterally induced due to the applied surface 
displacement, we find that it is transmitted into the material with an apparent phase difference.
We also obtain the effective complex modulus of the layered material and show that it exhibits an 
intermediate power-law behavior in the low-frequency regime. 
Our result can be used to estimate the internal deformation of layered materials that exist on various 
different scales. 
\end{abstract}

\maketitle

\section{Introduction}
\label{Sec:introduction}

Although the rheology of soft materials is an important subject, their nonlinear viscoelasticity as well as 
linear viscoelastic behaviors are not fully understood.
For a structurally homogeneous material, the linear viscoelasticity can be characterized by the frequency-dependent 
complex modulus $G^{\ast}(\omega)$. 
Typical soft materials, however, contain mesoscopic internal structures that can deform under weak external fields, 
leading to unique viscoelastic behaviors~\cite{LarsonBook,WittenBook}.
These internal structures play essential roles for the mechanical response at low-frequencies, whereas the molecular 
interactions control the high-frequency response.
To describe the internal mesoscopic structures of soft materials, a coupling mechanism between elastic and 
viscous components have been often considered.  
In the two-fluid model for a polymer gel, for instance, the consisting polymer network is represented by an 
elastic material and the solvent is described as a viscous fluid~\cite{deGennes76a,deGennes76b}.

The mechanical response of a layered material such as the smectic phase in liquid crystals~\cite{Colby97,Lu08,Fujii1,Fujii3,Fujii2,Fujii4} 
or the lamellar phase in diblock copolymers~\cite{Fredrickson96}. 
In the experiments, the orientation of the stacked layers can be controlled by various methods. 
For example, the lamellar structure in block copolymers can be oriented in the flow direction under a weak 
shear flow, while it becomes perpendicular to the flow for a strong shear flow~\cite{Fredrickson96}. 
To explain the anomalous frequency dependence of the complex modulus of the lamellar structure in block 
copolymers~\cite{Fredrickson96,Bates84,Rosedale90,Bates90}, Kawasaki and Onuki considered the coupling between the 
bending elasticity of the lamellar layers and the viscosity of the surrounding fluid~\cite{Kawasaki90}.

Layered structures exist not only in soft materials but also in biological materials, hard condensed matter, 
and in the field of geology.
The corresponding examples are epithelial tissues~\cite{The Cell}, graphene sheets~\cite{Kolahchi,Mehrez}, 
layered viscoelastic materials~\cite{Xu} and geological strata, respectively. 
However, a general model to describe such a layered structure has not yet been considered.

In this paper, we propose a continuum model for an oriented layered material that 
contains lateral elasticity and interlayer friction to investigate its mechanical properties. 
In our model, each layer is described as a two-dimensional (2D) elastic sheet that undergoes both shear 
and areal deformations.
The frictional interaction between the layers is phenomenologically introduced through the 
velocity gradient of the layers. 
Due to this coupling effect, the overall mechanical response becomes viscoelastic and frequency-dependent.
Considering only the in-plane deformation, we obtain the response function that relates 
the surface displacement to the internal deformation both in the Fourier space and real spaces. 
We examine how the surface lateral displacement is transmitted into the material as a function 
of the depth or the distance.  
Moreover, we shall obtain the effective complex modulus of the layered material and discuss its asymptotic
behaviors.

Before presenting our model, we shall clarify the differences between fluid membranes 
and elastic sheets.
Previously, both the static and dynamic properties of a stack of fluid membranes or the lamellar phase in lyotropic liquid crystals were studied experimentally and theoretically~\cite{Nallet89,Nallet94,Ramaswamy93}. 
In particular, the shear-induced transition from the lamellar phase to the onion phase in lyotropic systems was investigated~\cite{Roux93,Bonn98,Panizza,Zilman99,Marlow02}. 
The deformation of a fluid membrane is described by its out-of-plane displacement because it is incompressible and there is no restoring force against shear deformation. 
For an elastic sheet, however, the in-plane and out-of-plane displacements are coupled to each other in a non-linear manner and its deformation is highly nontrivial~\cite{Landau,Nelson87,Doussal}. 
This is one of the reasons that we limit our analysis only to the in-plane lateral response and further discussion concerning the extension of our work will be given in the last section.

For a fluid membrane which has two-dimensional shear viscosity, the inter-leaflet friction between two monolayers 
in a bilayer membrane is recognized as another source of dissipation~\cite{Seifert93,Okamoto16,Okamoto17,Yasuda18}. 
Such friction arises due to the velocity difference between the upper and lower leaflets of a the bilayer.
However, the interlayer friction between elastic sheets was not considered before, and we shall propose the simplest continuum model to take into account such an effect.

In Sec.~\ref{Sec:model}, we describe our continuum model for a layered material with interlayer friction.
In Secs.~\ref{Sec:FourierSpace} and \ref{Sec:RealSpace}, we obtain the response functions in the 
Fourier and the real spaces, respectively, and further discuss the longitudinal and transverse 
responses. 
The effective complex modulus of a layered material is calculated in Sec.~\ref{Sec:modulus}. 
Finally, a summary and some further discussion are given in Sec.~\ref{Sec:discussion}.

\section{Model of a layered material}
\label{Sec:model}

\begin{figure}[t]
\begin{center}
\includegraphics[scale=0.28]{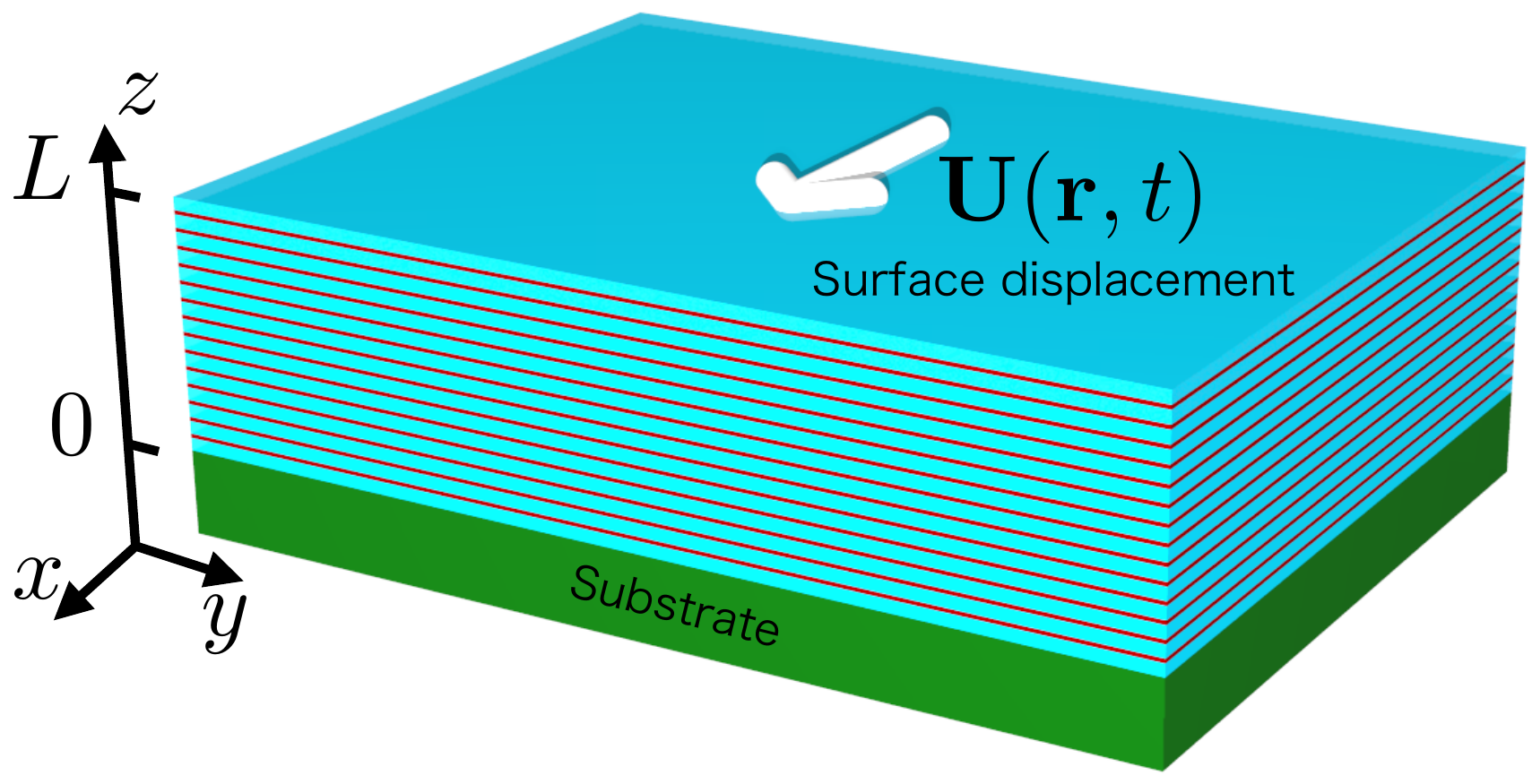}
\end{center}
\caption{Layered material of thickness $L$ consisting of two-dimensional elastic sheets (red layers) 
and thin fluid (blue layers). 
The surface displacement $\mathbf{U}(\mathbf r,t)$ at $z=L$  propagates into the material.
The layered material is supported by a solid substrate at $z=0$ where the deformation vanishes.
}
\label{LayeredElasticSolid}
\end{figure}

We first discuss the stress tensor of a 2D elastic sheet that comprises the layered material as 
shown in Fig.~\ref{LayeredElasticSolid} where the layers are stacked in the $z$-direction.
We introduce the time-dependent 2D lateral in-plane displacement field 
$\mathbf{u}(\mathbf{r},z,t)=(u_x(\mathbf{r},z,t),u_y(\mathbf{r},z,t))$ where $\mathbf{r}=(x,y)$
is the 2D coordinate.
In this work, we do not consider any out-of-plane deformation of the sheet.
Then the strain tensor can be defined by
\begin{align}
\varepsilon_{ij}=\frac{1}{2} \left( \partial_i u_j + \partial_j u_i \right), 
\label{strain_tensor}
\end{align}
where $i, j =x,y$.
Within the linear elasticity theory, the 2D stress tensor is related to the above strain tensor by the 
Hook's law~\cite{Landau} 
\begin{align}
\sigma_{ij} = \lambda \varepsilon_{kk} \delta_{ij} + 2 \mu\varepsilon_{ij},
\label{strain_stress_tensor}
\end{align}
where $\delta_{ij}$ is the Kronecker delta, $\lambda$ and $\mu$ are the 2D Lam\'e coefficients,
and the summation over repeated indices is implicitly assumed.
Here, the symmetry $\sigma_{ij} =\sigma_{ji}$ implies the 2D isotropy 
of the elastic sheet.

To characterize the layered structure, we consider frictional forces acting between the elastic sheets.
In the small displacement limit, the frictional force is proportional to the velocity difference between 
the two neighboring sheets.
When the thickness of each layer $h$ is small enough, the stress acting between them can be expressed 
by a derivative and the intra-layer stress becomes 
\begin{align}
\sigma_{iz}^{\mathrm {int}} = \frac{\zeta}{h} (\partial_z\dot{u}_i),
\label{Thin Sheet Stress}
\end{align}
where $\zeta$ is the friction coefficient and the dot indicates the time derivative.
The above frictional stress arises from, e.g., a viscous fluid between the sheets and 
can be obtained within the lubrication approximation, as shown in Appendix~\ref{App:friction}.
However, we note that Eq.~(\ref{Thin Sheet Stress}) is not limited to such a situation and holds
more generally for other layered materials.

Using the above result, we now obtain the equation of motion of a layered material with interlayer friction.
We assume that the thickness of each sheet is negligibly small.
By combining Eqs.~(\ref{strain_stress_tensor}) and (\ref{Thin Sheet Stress}), the equation of motion of the 
layered material is generally given by 
$\rho \ddot u_i=\partial_j\sigma_{ij}+h \partial_z \sigma_{iz}^{\mathrm {int}}$,
where $\rho$ is the average areal density.
If we neglect the effect of inertia under the condition $\rho L^2\omega/\zeta \ll1$, where $L$ is the 
thickness of the layered material, we obtain the following force balance equation:
\begin{align}
\mu \partial_j^2u_i + \left( \mu+\lambda \right) \partial_i\partial_j u_j + \zeta \partial_z^2\dot u_i=0.
\label{elastic eq.}
\end{align}

As shown in Fig.~\ref{LayeredElasticSolid}, the bottom of the layered material ($z=0$) is supported by a 
solid substrate and the displacement vanishes there.
On the other hand, we apply displacement $\mathbf{U}(\mathbf r,t)$ at the top surface ($z=L$).
Hence the boundary conditions for the displacement field are written as  
\begin{align}
\mathbf{u}(\mathbf r,z,t)=
\begin{cases}
0 &{\rm at}~z=0
\\
\mathbf{U}(\mathbf r,t) &{\rm at}~z=L
\end{cases}.
\label{boundary condition}
\end{align}

\section{Response function in the Fourier space}
\label{Sec:FourierSpace}

\begin{figure}[t]	
\begin{center}
\includegraphics[scale=0.5]{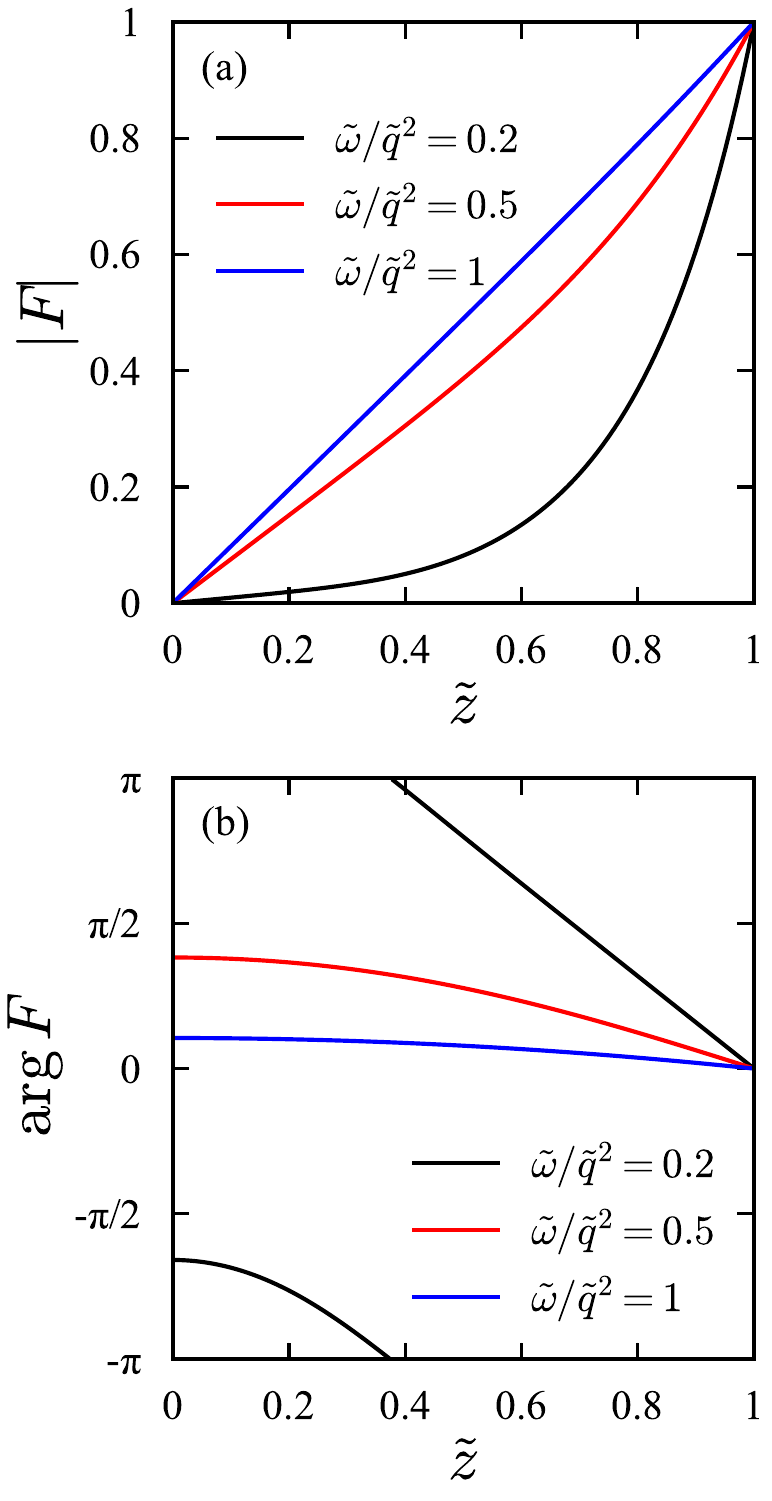}
\caption{The plots of (a) the absolute value and (b) the argument of the complex function 
$F[\tilde{q}, \tilde{z}, \tilde{\omega}]$ in Eq.~(\ref{Diff-R}) as a function of $\tilde z=z/L$ 
for $\tilde\omega/\tilde q^2=0.2$, $0.5$, and $1$.}
\label{FourierC1_Abs_Arg}
\end{center}
\end{figure}

In this section, we discuss the mechanical response of the layered material described by Eq.~(\ref{elastic eq.})
under the boundary conditions in Eq.~(\ref{boundary condition}).
We first introduce the 2D Fourier transform defined by
\begin{align}
u_i[\mathbf{q}, z, \omega]& = \int d^2 \mathbf{r} \int dt \, u_i(\mathbf{r}, z, t ) 
e^{-i (\mathbf{q} \cdot \mathbf{r}+ \omega t)}, 
\label{Fourier transform eq.}\\
u_i(\mathbf{r},z,t) & = \int \frac{d^2 \mathbf{q}}{(2 \pi)^2} \int \frac{d \omega}{2 \pi} \,u_i[\mathbf{q}, z, \omega]
e^{i (\mathbf{q} \cdot \mathbf{r} + \omega t)},
\label{inverse-Fourier transform eq.}
\end{align}
where $\mathbf{q}=(q_x,q_y)$ is the 2D wavevector and $\omega$ is the frequency. 
Then the Fourier transform of Eq.~(\ref{elastic eq.}) becomes 
\begin{align}
\partial_z^2 u_i[\mathbf{q}, z, \omega] = A_{ij}[\mathbf{q}, \omega] u_j [\mathbf{q}, z, \omega],
\label{Fourier-elastic eq.} 
\end{align}
where we have introduced 
\begin{align}
A_{ij}[\mathbf{q}, \omega]=
\frac{q^2}{i \omega \zeta} 
\left[ \mu \delta _{ij} + (\mu + \lambda)\frac{q_iq_j}{q^2} \right],
\end{align}
and $q=|\mathbf{q}|$.

Equation~(\ref{Fourier-elastic eq.}) is a second-order linear differential equation that can be solved easily.  
The eigenvalues $\xi_\pm$ and the corresponding eigenvectors  $\mathbf w_\pm$ of the matrix $A_{ij}$ 
are obtained as 
\begin{align}
\xi_{+} & = \frac{q^2}{i \omega \zeta} \left(2 \mu +  \lambda \right), \quad  
\mathbf w_{+} 
= \left( 
\begin{array}{c}
q_x \\
q_y \\
\end{array}
\right), \\
\xi_{-} & = \frac{q^2}{i \omega \zeta}  \mu,  \quad \quad \quad \quad 
\mathbf w_{-} 
= \left( 
\begin{array}{c}
q_y \\
- q_x \\
\end{array}
\right).
\end{align}
Then the general solution of $u_i[\mathbf q, z,\omega]$ is obtained as 
\begin{align}
\mathbf{u}[\mathbf{q}, z, \omega]
= \left( 
\begin{array}{cc}
    q_x   &  q_y \\
    q_y   &  -q_x \\
   \end{array}
\right)
\left( 
\begin{array}{cc}
   a_1 e^{\sqrt{\xi_{+}} z}   +  a_2 e^{- \sqrt{\xi_{+}} z} \\
   a_3 e^{\sqrt{\xi_{-}} z}   +  a_4 e^{- \sqrt{\xi_{-}} z} \\
 \end{array}
\right).
\label{general solution}
\end{align}
In the above, $a_1, \dots, a_4$ are the coefficients determined by the boundary conditions in 
Eq.~(\ref{boundary condition}) and are given by 
\begin{align}
a_1 & = -a_2 = \displaystyle \frac{U_x q_x + U_y q_y}{2q \sinh \left(\sqrt{\xi_{+}} L \right)}, \\
a_3 & = -a_4 = \displaystyle \frac{U_x q_y - U_y q_x}{2q \sinh \left(\sqrt{\xi_{-}} L \right)}. 
\end{align}

The linear response to the applied displacement $\mathbf{U}(\mathbf r,t)$ at the materials surface ($z=L$) 
can be written in the following way:
\begin{align}
u_i[\mathbf{q}, z, \omega] =\gamma_{ij}[\mathbf{q}, z, \omega] U_j[\mathbf{q},\omega].
\label{Displacement in Fourier space}
\end{align}
Here, $\gamma_{ij}[\mathbf{q}, z, \omega]$ is the response function in the Fourier space and can be
represented as 
\begin{align}
\gamma_{ij}[\mathbf{q}, z, \omega] =B_1[\mathbf{q}, z, \omega] \delta_{ij} + B_2[\mathbf{q}, z, \omega] \frac{q_i q_j}{q^2}.
\label{Fourier_G_def}
\end{align}
After some calculation, we obtain 
\begin{align}
B_1 [\mathbf{q}, z, \omega] & =F[\tilde{q}, \tilde{z}, \tilde{\omega}_\perp],
\\ 
B_2 [\mathbf{q}, z, \omega] & =F[\tilde{q}, \tilde{z}, \tilde{\omega}_\parallel]  - F[\tilde{q}, \tilde{z}, \tilde{\omega}_\perp],
\end{align}
where the function $F[\tilde{q}, \tilde{z}, \tilde{\omega}]$ is defined by 
\begin{align}
F[\tilde{q}, \tilde{z}, \tilde{\omega}]=\frac{\sinh \left[ (1-i)\tilde q \tilde z/ \sqrt{\tilde{\omega}}\right]}
{\sinh \left[  (1-i)\tilde q/\sqrt{\tilde{\omega}} \right]},
\label{Diff-R}
\end{align}
and the dimensionless quantities are $\tilde q = qL$, $\tilde z = z/L$, 
$\tilde{\omega}_\perp = (2 \zeta/\mu) \omega$, and 
$\tilde{\omega}_\parallel = [2 \zeta/(2\mu + \lambda)] \omega$.
Note that $\tilde{\omega}_\parallel$ and $\tilde{\omega}_\perp$ are related by 
$\tilde{\omega}_\parallel=(1-\nu)\tilde{\omega}_\perp/2$, where $\nu=\lambda/(2\mu+\lambda)$ 
is the 2D Poisson ratio.
In Eq.~(\ref{Diff-R}), the frequency dependence enters only with the combination $\tilde \omega/\tilde q^2$.

Using the obtained response function, we discuss the internal displacement of the layered material.
Owing to the 2D isotropy, we set here $U_y=0$ for simplicity and without loss of generality.
If we apply a transverse displacement $U_x \sim e^{i (qy + \omega t)}$ at $z=L$, the internal 
displacement is given by $u_x[\mathbf{q}, z, \omega]=B_1=F[\tilde{q}, \tilde{z}, \tilde{\omega}_\perp]$ 
according to Eqs.~(\ref{Displacement in Fourier space}) and (\ref{Fourier_G_def}). 
For a longitudinal displacement $U_x \sim e^{i (qx + \omega t)}$ at $z=L$, on the other hand, 
the internal displacement is given by 
$u_x[\mathbf{q}, z, \omega]=B_1+B_2=F[\tilde{q}, \tilde{z}, \tilde{\omega}_\parallel]$.
Hence the internal displacement is essentially expressed by the function $F[\tilde q, \tilde{z}, \tilde{\omega}]$
in Eq.~(\ref{Diff-R}) both for the transverse and longitudinal surface displacements at $z=L$.

Let us look at the asymptotic behaviors of the function $F[\tilde{q}, \tilde{z}, \tilde{\omega}]$.
In the limit of $\tilde\omega\gg\tilde q^2$, Eq.~(\ref{Diff-R}) becomes
\begin{align}
F[\tilde q, \tilde{z}, \tilde{\omega}]\approx\tilde z.
\end{align}
This means that the displacement is linearly transmitted through the layered material.
In the limit of $\tilde\omega\ll\tilde q^2$ and $\tilde z \approx 1$, we have 
\begin{align}
F[\tilde q, \tilde{z}, \tilde{\omega}]&\approx\exp \left[ (1-i)\tilde q (\tilde z-1)/ \sqrt{\tilde{\omega}}\right].
\label{R-large-q}
\end{align}
This expression indicates that $\sqrt{\tilde{\omega}}/\tilde q$ corresponds to the screening length 
beyond which the displacement decays out.

In Fig.~\ref{FourierC1_Abs_Arg}, we plot the function $F[\tilde{q}, \tilde{z}, \tilde{\omega}]$ 
as a function of $\tilde{z}$ for $\tilde{\omega}/\tilde q^2=0.2$, $0.5$, and $1$. 
Since $F$ is a complex quantity and can be written as $F=|F|e^{i(\mathrm{arg}\,F)}$, 
we plot its absolute value $|F|$ and the argument $(\mathrm{arg}\,F)$ in (a) and (b), respectively. 
In Fig.~\ref{FourierC1_Abs_Arg}(a), we see that $\vert F \vert$ interpolates between $0$ and $1$
in a nonlinear manner, and it is smaller when $\tilde{\omega}/\tilde q^2$ is smaller. 
This means that the displacement at $z=L$ cannot be transmitted deep into the layered material when 
the frequency is small.
In Fig.~\ref{FourierC1_Abs_Arg}(b), we see that the transmission of the displacement is delayed when 
$\tilde{\omega}/\tilde q^2$ is smaller. 
This is because each sheet undergoes different displacements with different phases when the 
frequency is small.
When the argument of $F$ is close to $\pi$ or $-\pi$, the phase is completely opposite to the applied 
displacement at $z=L$.

\begin{figure}[t]
\includegraphics[scale=0.5]{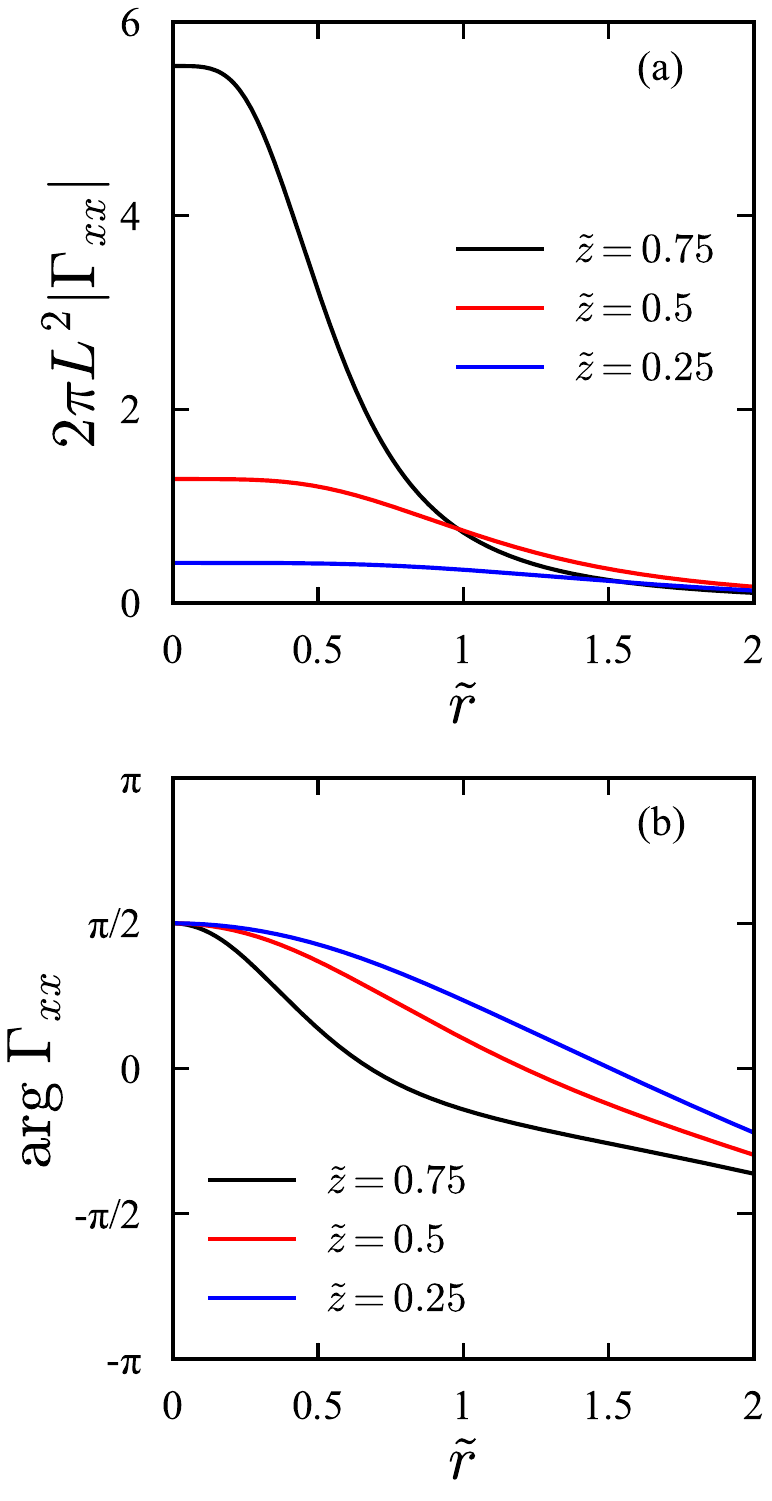}
\caption{The plots of (a) the absolute value and (b) the argument of the complex function 
$\Gamma_{xx}(\mathbf{r}, z, \omega)$ in Eq.~(\ref{real_G_def}) as a function of $\tilde r=r/L$ 
for $\tilde{z}=0.25$, $0.5$, and $0.75$. 
The other parameters are $\tilde{\omega}_\perp=1$ and $\nu=1/6$. 
}
\label{Gxx(r)_o1_Abs_Arg}
\end{figure}

\begin{figure}[t]
\includegraphics[scale=0.5]{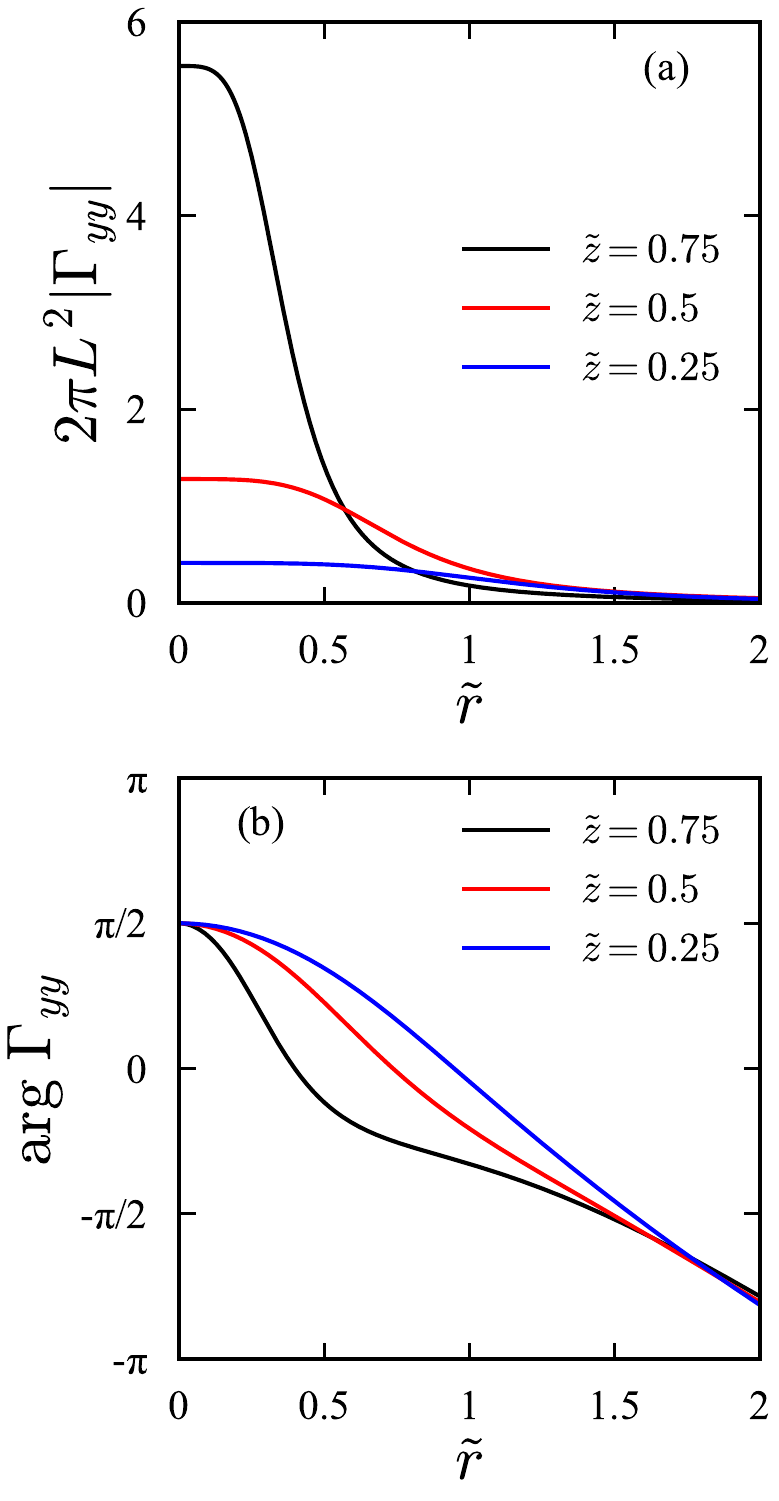}
\caption{The plots of (a) the absolute value and (b) the argument of the complex function 
$\Gamma_{yy}(\mathbf{r}, z, \omega)$ in Eq.~(\ref{real_G_def}) as a function of $\tilde r=r/L$ 
for $\tilde{z}=0.25$, $0.5$, and $0.75$. 
The other parameters are $\tilde{\omega}_\perp=1$ and $\nu=1/6$. 
}
\label{Gyy(r)_o1_Abs_Arg}
\end{figure}

\section{Response function in the real space}
\label{Sec:RealSpace}

Using the Fourier space response function in Eq.~(\ref{Fourier_G_def}), we now obtain the 
response function in the real space. 
In the real space, Eq.~(\ref{Displacement in Fourier space}) can be written as 
\begin{align}
u_i(\mathbf{r}, z,\omega) =\int d^2 \mathbf{r'} \, \Gamma_{ij}(\mathbf{r - r'}, z, \omega) U_j(\mathbf{r'}, \omega ), 
\end{align}
where 
\begin{align}
\Gamma_{ij}(\mathbf{r}, z, \omega) = \int \frac{d^2 \mathbf{q}}{(2 \pi)^2} \, 
\gamma_{ij}[\mathbf{q}, z, \omega]e^{i \mathbf{q\cdot r}}.
\label{real_G_integrate}
\end{align}

Due to the symmetry property $\Gamma_{ij}=\Gamma_{ji}$, we can write the response function in the form 
\begin{align}
\Gamma_{ij}(\mathbf{r}, z, \omega)= C_1(r, z, \omega) \delta_{ij} 
+ C_2(r, z, \omega) \frac{r_i r_j}{r^2},
\label{real_G_def}  
\end{align}
where $r=|\mathbf{r}|$.
In the above, $C_1$ and $C_2$ are obtained by calculating $\Gamma_{ii}$ and $\Gamma_{ij}r_ir_j/r^2$.
From Eqs.~(\ref{Fourier_G_def}) and (\ref{real_G_integrate}), the integral forms of $C_1$ and $C_2$ are  
\begin{align}
&C_1(\tilde{r},\tilde{z},\tilde{\omega})= \frac{1}{2 \pi L^2} \int^{\infty}_{0} d\tilde q \, \tilde q \bigg[ J_0(\tilde q \tilde{r}) 
B_1[\tilde q ,\tilde{z},\tilde{\omega}]
\nonumber \\
&+ \left(J_0(\tilde q \tilde{r}) - \frac{J_{1}(\tilde q \tilde{r})}{\tilde q \tilde{r}} + J_{2}(\tilde q \tilde{r}) \right) B_2[\tilde q,\tilde{z},\tilde{\omega}] \bigg], 
\label{Real_Green_Function_A}
\end{align}
\begin{align}
C_2(\tilde{r},\tilde{z},\tilde{\omega}) &= \frac{1}{ 2\pi L^2}\int^{\infty}_{0} d\tilde q \, \tilde q \bigg[- J_0(\tilde q \tilde{r})
\nonumber \\
&+\frac{2J_{1}(\tilde q \tilde{r})}{\tilde q \tilde{r}} -  2J_{2}(\tilde q \tilde{r})  \bigg] B_2[\tilde q,\tilde{z},\tilde{\omega}],
\label{Real_Green_Function_B}
\end{align}
where $J_n(x)$ is the Bessel function of the first kind and $\tilde r=r/L$.
These integrals can be numerically performed by using Mathematica.

Having obtained the response function $\Gamma_{ij}$ in the real space, we discuss here its longitudinal and 
transverse components.
Given the surface deformation $\mathbf{U}(\mathbf r, \omega)=\mathbf{U}(\omega) \delta(\mathbf r)$ at $z=L$, 
we obtain the internal displacement $\mathbf{u}(r, z, \omega)$.
If we choose the $x$-axis to be the direction of the applied surface displacement at $z=L$, one can consider the 
longitudinal response $\Gamma_{xx}=C_1+C_2$ and the transverse response $\Gamma_{yy}=C_1$ as before.

In Figs.~\ref{Gxx(r)_o1_Abs_Arg}(a) and (b), we plot the absolute value and the argument of complex 
$\Gamma_{xx}$, respectively, as a function of $\tilde r$ for $\tilde{z}=0.75$, $0.5$, and $0.25$. 
We see that $\vert \Gamma_{xx} \vert$ monotonically decreases as $\tilde r$ increases. 
More precisely, $\vert \Gamma_{xx} \vert$ decreases as $- r^2$ for $\tilde r\ll1$ while
it decays as $r^{-3}$ for $\tilde r\gg1$, as derived in Appendix~\ref{App:Approx}.
In Fig.~\ref{Gxx(r)_o1_Abs_Arg}(b), we see that the argument of $\Gamma_{xx}$ is $\pi/2$ at $\tilde{r}=0$, 
indicating that the displacement at $z=L$ is transmitted through the interlayer friction for $\tilde{r}=0$.
When $\tilde r$ is further increased, the argument of $\Gamma_{xx}$ monotonically decreases even 
to negative values.
In Figs.~\ref{Gyy(r)_o1_Abs_Arg}(a) and (b), we plot the absolute value and the argument of 
$\Gamma_{yy}$, respectively, as a function of $\tilde r$ for $\tilde{z}=0.75$, $0.5$, and $0.25$. 
Although the behavior of $\Gamma_{yy}$ is similar to that of $\Gamma_{xx}$,  the transverse response
$\Gamma_{yy}$ decays slightly faster than $\Gamma_{xx}$.

\section{Effective complex modulus}
\label{Sec:modulus}

\begin{figure}[t]
\begin{center}
\includegraphics[scale=0.35]{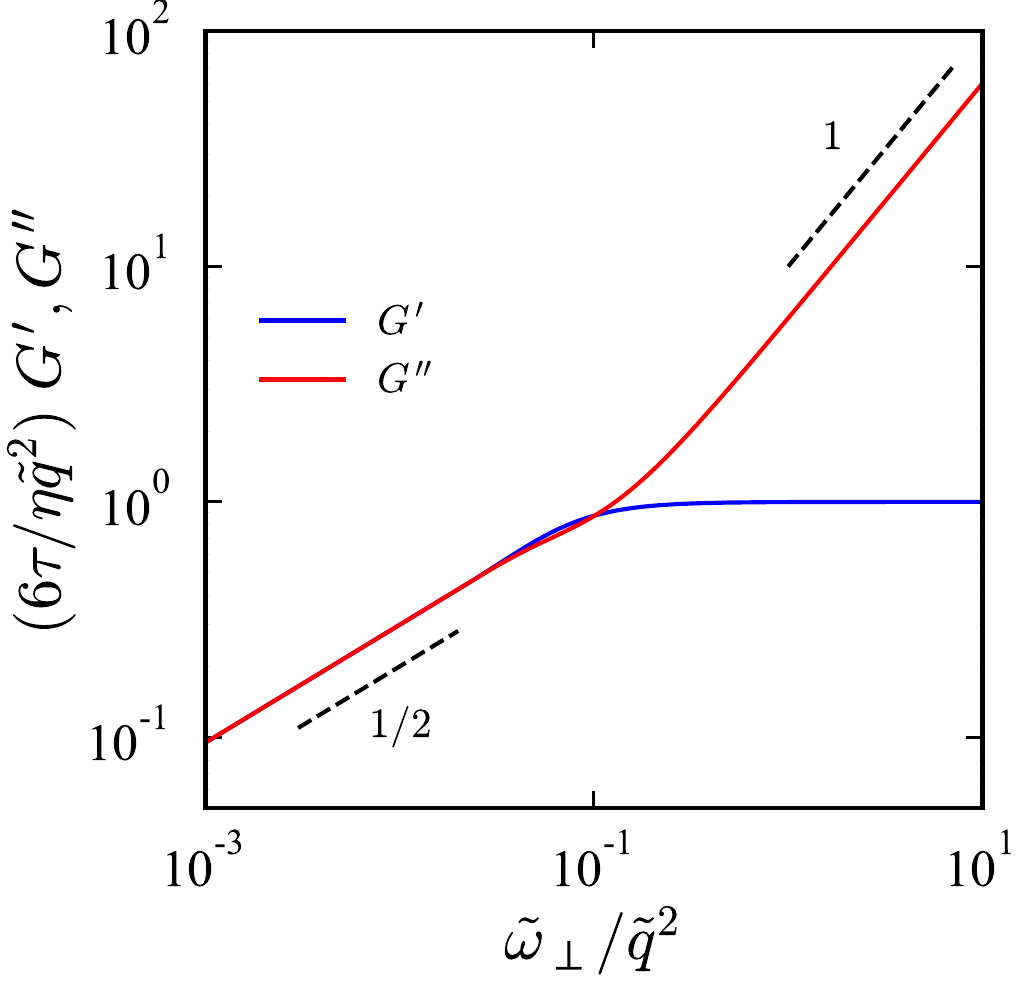}
\end{center}
\caption{The real part ($G'$) and the imaginary part ($G''$) of the effective complex modulus 
$G^{\ast}[q, \omega]$ in Eq.~(\ref{effmodulus}). 
The dashed lines represent the asymptotic power-law behaviors in Eqs.~(\ref{G_Asympto_L}) and (\ref{G_Asympto_S}).}
\label{Complex Elasticity}
\end{figure}

For the layered material shown in Fig.~\ref{LayeredElasticSolid}, we define an effective complex modulus
$G^{\ast}[\mathbf q, \omega]$ by using the stress $\sigma_{xz}^L$ at $z=L$ and $\sigma_{xz}^0$ at $z=0$ as 
\begin{align}
\frac{1}{2}\left( \sigma_{xz}^L[\mathbf q, \omega]+ \sigma_{xz}^0[\mathbf q, \omega]\right)
=\frac{1}{L} G^{\ast}[\mathbf q, \omega] U_x[\mathbf q,\omega].
\end{align}
Here we have set $U_y=0$ and use Eqs.~(\ref{Thin Sheet Stress}) and (\ref{Displacement in Fourier space}) 
to obtain 
\begin{align}
\sigma_{xz}^{L,0}[\mathbf q, \omega]=\frac{i\omega\zeta}{h} \partial_z\gamma_{xx}[\mathbf q,z,\omega]|_{z={L,0}} 
U_x[\mathbf q,\omega]. 
\end{align}
Since $\gamma_{xx}$ in Eq.~(\ref{Fourier_G_def}) is essentially given by $F[\tilde q, \tilde{z}, \tilde{\omega}_\perp]$
in Eq.~(\ref{Diff-R}) for the transverse surface displacement, the effective modulus $G^{\ast}[\mathbf q, \omega]$ 
is obtained as 
\begin{align}
& G^{\ast}[\mathbf q, \omega] =\frac{i\omega\zeta L}{2h}
\left(\partial_z F[\tilde{q}, \tilde{z}, \tilde{\omega}_\perp]|_{z=L}+\partial_z F[\tilde{q}, \tilde{z}, \tilde{\omega}_\perp]|_{z=0} \right)
\nonumber \\
&=\frac{i\omega\zeta(1-i)\tilde q}{2h\sqrt{\tilde{\omega}_\perp}}
\nonumber \\
& \times \left[\frac{1}{\tanh \left[  (1-i)\tilde q/\sqrt{\tilde{\omega}_\perp} \right]}
+\frac{1}{\sinh \left[  (1-i)\tilde q/\sqrt{\tilde{\omega}_\perp} \right]}\right].
\label{effmodulus}
\end{align}
The behaviors of the real and imaginary parts of $G^{\ast}$ denoted by $G'$ and $G''$, respectively, 
are shown in Fig.~\ref{Complex Elasticity}.

Let us discuss the asymptotic behaviors of Eq.~(\ref{effmodulus}).
In the limit of $\tilde\omega_\perp\gg\tilde q^2$, we have 
\begin{align}
G^{\ast}[q, \omega] \approx \frac{i\omega\zeta}{h} + \frac{\zeta \tilde q^2}{6h\tau}, 
\label{G_Asympto_L}
\end{align}
where $\tau=2\zeta/\mu$ is the relaxation time for the transverse surface displacement 
as we discussed in Sec.~\ref{Sec:FourierSpace}.
The above expression is analogous to the complex modulus of the Kelvin-Voigt model. 
In the limit of $\tilde\omega_\perp\ll\tilde q^2$, on the other hand, we have 
\begin{align}
G^{\ast}[q, \omega] \approx \frac{\sqrt{2}\zeta\tilde q}{2h\tau}(i\tilde\omega_\perp)^{1/2},  
\label{G_Asympto_S}
\end{align}
which is an intermediate power-law behavior. 
In the long-length limit, however, the layered material behaves as a fluid because 
$G^{\ast} \to i\omega\zeta/h$ for $q\to0$.

It was experimentally found that the complex modulus of the lamellar phase in diblock copolymers 
behaves as $G^{\ast} \sim (i \omega)^{1/2}$~\cite{Fredrickson96,Bates84,Rosedale90,Bates90}. 
Using the smectic free energy of liquid crystals, Kawasaki and Onuki proposed a theory to describe 
the intermediate behavior between fluids and solids~\cite{Kawasaki90}.
Although we obtain a similar frequency dependence in Eq.~(\ref{G_Asympto_S}), we do not consider that 
our result explains the experimental observation in block copolymers because of the following reasons.
(i) In the experiment, the anomalous behavior $G^{\ast} \sim (i \omega)^{1/2}$ disappears if layers are oriented 
by application of shear.
(ii) The power-law behavior in Eq.~(\ref{G_Asympto_S}) appears only for finite $q$ and cannot be a macroscopic 
response for $q\to0$.

\section{Summary and discussion}
\label{Sec:discussion}

In this paper, we have discussed the mechanical properties of a layered material with interlayer friction.
We have proposed a new model that contains both lateral elasticity and interlayer friction.
Considering only the in-plane deformation, we obtained the response function in the Fourier and 
the real spaces. 
In particular, we have looked at how the internal displacement is induced by the surface lateral displacement.
We find that the applied surface displacement is transmitted into the material with an apparent phase 
difference due to the interlayer friction.
We also obtained the effective complex modulus of the layered material and showed that it exhibits an 
intermediate power-law behavior in the low-frequency regime. 
Our model is general and can be applied for various layered materials at different scales ranging 
from microscopic to macroscopic sizes.

In our model, we have assumed that the 2D sheets are elastic and interlayer friction acts between them.
As shown in Appendix~\ref{App:friction}, such a friction can be obtained, for example, within the lubrication 
approximation for the fluid layer between the elastic sheets.
It is also similar to that acting between two leaflets in a fluid bilayer membrane~\cite{Seifert93,Okamoto16,Okamoto17, Yasuda18}.
On the other hand, the situation is different in graphene sheets~\cite{Kolahchi,Mehrez} or layered viscoelastic materials~\cite{Xu} 
in which elastic interactions act between the sheets.
Since elastic interactions do not lead to any dissipation, the mechanical response should be different from 
what we have discussed in the paper.

In the present work, we have discussed only the lateral in-plane displacement at the surface of the layered material 
and investigated how it is transmitted into the material.
In the future, we shall also investigate the mechanical response to the out-of-plane surface displacement.
For a thin elastic sheet, it is known that in-plane and out-of-plane displacements are coupled to each other so that bending 
always accompanies stretching~\cite{Landau}.  
For such a non-linear coupling, one can eliminate the in-plane degrees of freedom, which results in the renormalization 
of the bending rigidity~\cite{Nelson87,Doussal}. 
A hydrodynamic theory for a single polymerized membrane was discussed in Ref.~\cite{Frey91} by focusing on the dynamics 
of out-of-plane deformation that is coupled to the surrounding fluid.

In order to consider the mechanical response to the out-of-plane deformation in the layered material, one needs to further 
take into account such as the bending rigidity and layer compression modulus that are used to describe the elasticity of 
the smectic phase in liquid crystals~\cite{deGennesBook}. 
For multi-layered elastic sheets, the steric interactions between two elastic sheets are very different from those between fluid membranes~\cite{Leibler89}.
Moreover, it is necessary to investigate the effects of hydrodynamic interactions between elastic sheets for out-of-plane deformation. 
These extended studies are left for our future work. 
However, we emphasize that even the lateral response to in-plane deformation appears to be non-trivial due to the friction 
between the layers, as we have discussed in this paper.

In our previous work on two-layer vesicles~\cite{Lu12}, we argued both the bending energy and the 
stretching energy of the fluid membranes as well as the hydrodynamics of the surrounding 
fluid. 
We can also extend this theory to a multi-layer system by including the shear elasticity of each sheet.

\begin{acknowledgments}

K.Y.\ acknowledges the support by a Grant-in-Aid for JSPS Fellows (Grant No.\ 21J00096) from the 
Japan Society for the Promotion of Science.
K.Y.\ was supported by the Research Institute for Mathematical Sciences, an International 
Joint Usage/Research Center located in Kyoto University.
S.K.\ acknowledges the supported by the startup fund of Wenzhou Institute, University of Chinese Academy 
of Sciences (No.\ WIUCASQD2021041). 
\end{acknowledgments}

\appendix
\section{The interlayer friction due to viscous fluid}
\label{App:friction}

In this Appendix, we derive Eq.~(\ref{Thin Sheet Stress}) for the situation in which a 3D viscous fluid 
exists between the elastic sheets.
The stress tensor of a 3D incompressible fluid is given by~\cite{Landau_Fluid} 
\begin{align}
\sigma_{\alpha\beta}^{\mathrm {3D}} = -p \delta_{\alpha\beta} + \eta \left(\partial_\alpha v_\beta + 
\partial_\beta v_\alpha \right),
\label{viscosity stress}
\end{align}
where $p$ is the 3D pressure, $\mathbf{v}$ is the 3D fluid velocity, $\eta$ is the 3D shear viscosity, and 
$\alpha, \beta =x,y,z$.
Moreover, we assume that the fluid satisfies the incompressibility condition, $\partial_\alpha v_\alpha=0$.

To consider the stress acting on each sheet, we apply the lubrication approximation~\cite{Safran94}. 
We denote the thickness of the lubrication layer by $h$ and the velocity at the upper plane by $V_x$,
whereas the velocity on the elastic sheet vanishes. 
Then the hydrodynamic equations for the thin fluid layer are 
\begin{align}
\partial_zp=0,~~~\eta\partial_z^2 v_x-\partial_xp=0,~~~\partial_x v_x+\partial_z v_z=0.
\end{align}
In the absence of the pressure gradient $\partial_x p$, the velocity profile is simply given by a Couette flow, 
i.e., $v_x = V_x z/h$.
We regard the average stress between the upper and the bottom planes as the intra-layer stress given by 
$\sigma_{xz}^{\mathrm {int}}=(\sigma_{xz}^h+\sigma_{xz}^0)/2=\eta V_x/h$.
When $h$ is small enough, the stress can be expressed by a derivative as in Eq.~(\ref{Thin Sheet Stress}),
where $\zeta/h$ should be identified as $\eta$.

\section{Asymptotic expressions of Eqs.~(\ref{Real_Green_Function_A}) and (\ref{Real_Green_Function_B})}
\label{App:Approx}

In this Appendix, we show the derivation of the asymptotic expressions of $C_1$ and $C_2$ 
in Eqs.~(\ref{Real_Green_Function_A}) and (\ref{Real_Green_Function_B}), respectively.
For $\tilde r=0$, the integrals in Eqs.~(\ref{Real_Green_Function_A}) and (\ref{Real_Green_Function_B}) 
can be performed analytically, and we obtain 
\begin{align}
C_1(\tilde{r}=0,\tilde{z},\tilde{\omega}) & = \frac{1}{2 \pi L^2}\frac{i(\tilde \omega_\perp+\tilde \omega_\parallel)}{16}
\nonumber \\
& \times \left[\psi'\left(\frac{1-\tilde z}{2}\right)-\psi'\left(\frac{1+\tilde z}{2}\right)\right], 
\end{align}
where $\psi'(x)=d\psi(x)/dx$ with $\psi(x)=\Gamma'(x)/\Gamma(x)$ and $\Gamma(x)$ is the Gamma function.
On the other hand, we have $C_2(\tilde{r}=0,\tilde{z},\tilde{\omega})=0$.

Next, we consider the limit of $\tilde r\ll1$ and the expansion of the Bessel functions around zero to examine 
the terms of order $\tilde r^2$.
Then the integrals in Eqs.~(\ref{Real_Green_Function_A}) and (\ref{Real_Green_Function_B}) can be performed 
analytically as 
\begin{align}
C_1(\tilde r,\tilde{z},\tilde{\omega})& \approx C_1(0,\tilde{z},\tilde{\omega})+\frac{1}{2 \pi L^2}\frac{3\tilde r^2}{32\cdot16}
(3\tilde\omega_\perp^2+\tilde\omega_\parallel^2)
\nonumber \\
& \times \left[\zeta_4\left(\frac{1-\tilde z}{2}\right)-\zeta_4\left(\frac{1+\tilde z}{2}\right)\right],
\end{align}
\begin{align}
C_2(\tilde r,\tilde{z},\tilde{\omega}]&\approx -\frac{1}{2 \pi L^2}\frac{3\tilde r^2}{32\cdot8}
(\tilde\omega_\perp^2-\tilde\omega_\parallel^2)
\nonumber \\
& \times \left[\zeta_4\left(\frac{1-\tilde z}{2}\right)-\zeta_4\left(\frac{1+\tilde z}{2}\right)\right],
\end{align}
where $\zeta_n(x)$ is the generalized Riemann zeta function defined as
\begin{align}
\zeta_n(x)=\sum_{k=0}^\infty(k+x)^{-n}.
\end{align}

Next, we discuss the case of $\tilde r\gg1$.
In this case, the Bessel functions in Eqs.~(\ref{Real_Green_Function_A}) and (\ref{Real_Green_Function_B}) can be expressed 
as $J_0(x)\sim x^{-1/2}\cos(x-\pi/4)$ but an analytical treatment is impossible. 
Here we use the approximation in Eq.~(\ref{R-large-q}) to obtain 
\begin{align}
&C_1(\tilde r,\tilde{z},\tilde{\omega})\approx\frac{1}{2 \pi L^2}\frac{3\sqrt{2}(1-i)(1-\tilde{z})}{4\tilde r^3 }
\frac{1}{\sqrt{\tilde\omega_\perp}},
\\
&C_2(\tilde r,\tilde{z},\tilde{\omega})\approx \frac{1}{2 \pi L^2}\frac{3\sqrt{2}(1-i)(1-\tilde{z})}{4\tilde r^3}
\left[\frac{1}{\sqrt{\tilde\omega_\parallel}}-\frac{1}{\sqrt{\tilde\omega_\perp}}\right].
\end{align}
In the above, we have used the relation
\begin{align}
\lim_{x\to\infty}[\cos(3\tan^{-1}(x)/2)+\sin(3\tan^{-1}(x)/2)]=\frac{3}{\sqrt{2}x}.
\end{align}
Hence we see that the response function decays as $r^{-3}$ at long distances.


\end{document}